\documentclass[preprint]{aastex61}

\shorttitle{The Threatening Environment of the TRAPPIST-1 Planets}

\shortauthors{Garraffo et al.}

\begin{document}

%\title{TRAPPIST-1 POSES A BIG THREAT TO ITS PLANETS' ATMOSPHERES}
\title{The Threatening Magnetic and Plasma Environment of the TRAPPIST-1 Planets}

\author{Cecilia Garraffo}
\affiliation{Harvard-Smithsonian Center for Astrophysics \\
60 Garden St., Cambridge, MA 02138, USA}

\author{Jeremy J. Drake}
\affiliation{Harvard-Smithsonian Center for Astrophysics \\
60 Garden St., Cambridge, MA 02138, USA}

\author{Ofer Cohen}
\affiliation{Lowell Center for Space Science and Technology, University of Massachusetts Lowell \\
600 Suffolk St., Lowell, MA 01854, USA}
\affiliation{Harvard-Smithsonian Center for Astrophysics \\
60 Garden St., Cambridge, MA 02138, USA}

\author{Julian D. Alvarado-G\'omez}
\affiliation{Harvard-Smithsonian Center for Astrophysics \\
60 Garden St., Cambridge, MA 02138, USA}

\author{Sofia P. Moschou}
\affiliation{Harvard-Smithsonian Center for Astrophysics \\
60 Garden St., Cambridge, MA 02138, USA}

\begin{abstract}

Recently, four additional Earth-mass planets were discovered orbiting the nearby ultracool M8 dwarf TRAPPIST-1, making a remarkable total of seven planets with equilibrium temperatures compatible with the presence of liquid water on their surface.  Temperate terrestrial planets around an M-dwarf orbit close to their parent star, rendering their atmospheres vulnerable to erosion by the stellar wind and energetic electromagnetic and particle radiation. Here, we use state-of-the-art 3D magnetohydrodynamic models to simulate the wind around TRAPPIST-1 and study the conditions at each planetary orbit.
All planets experience a stellar wind pressure between $10^3$ and $10^5$ times the solar wind pressure on Earth.  All orbits pass through wind pressure changes of an order of magnitude and most planets spend a large fraction of their orbital period in the sub-Alfv\'enic regime.  For plausible planetary magnetic field strengths, all magnetospheres are greatly compressed and undergo much more dynamic change than that of the Earth. The planetary magnetic fields connect with the stellar radial field over much of the planetary surface, allowing direct flow of stellar wind particles onto the planetary atmosphere. These conditions could result in strong atmospheric stripping and evaporation and should be taken into account for any realistic assessment of the evolution and habitability of the TRAPPIST-1 planets.

\end{abstract}

\footnote{email: cgarraffo@cfa.harvard.edu}

\section{Introduction} 

The recent discovery of four additional planets around TRAPPIST-1 with masses and radii similar to the Earth's \citep{Gillon.etal:17}, combined with the three already known \citep{Gillon.etal:16}, makes this system of special importance for characterizing terrestrial exoplanetary atmospheres and their evolution.

TRAPPIST-1 is an ultracool dwarf (M8V) 12~pc from the Sun, with a mass of $\sim 0.08~M_{\odot}$ and a radius of $R_\star= 0.114~R_{\odot}$. Its seven confirmed planets are in a co-planar system viewed nearly edge-on. All reside very close to the host star at distances from 0.01~AU to 0.063~AU (for comparison, Mercury is at 0.39~AU), with orbital periods from 1.5~days to 20~days. 
%, all with masses and radii similar to the Earth's. 

The atmospheres of close-in exoplanets are vulnerable to strong energetic (UV to X-ray) radiation and intense stellar wind conditions that could lead to atmospheric stripping \citep{Lammer.etal:03, Cohen.etal:14, Cohen.etal:15, Garraffo.etal:16}.  This is particularly important for planets in the habitable zones of M dwarfs whose low bolometric luminosities mean that temperate orbits lie very close to the parent star.  Even a substantial underestimation of the actual XUV emission of TRAPPIST-1 suggests that planets b and c could have lost up to 15 Earth oceans, while up to one Earth ocean could have escaped from planet~d \citep{Bolmont.etal:17,Wheatley.etal:17}. Climate models suggest that planet e represents the best chance for the presence of liquid water on its surface \citep{Wolf:17}. The capacity of these planets to have retained any water at all will depend critically on the initial water reservoir and on the erosive action of the stellar wind.

Stellar magnetic activity responsible for UV--X-ray emission and the generation of hot, magnetized winds in Sun-like stars is driven largely by stellar rotation. This influence extends from early F spectral types down to M9 and stars in the fully convective regime \citep{Wright.etal:11,Wright.Drake:16}. Activity increases with rotation rate up to a saturation limit beyond which faster rotation no longer results in further increase. Activity also appears to decline in very low mass stars and brown dwarfs with spectral types later than M9 \citep{Berger.etal:10}. $H_{\alpha}$ observations of 
TRAPPIST-1 \citep{Reiners.Basri:10} put its magnetic activity as high as the saturation regime, consistent with its M8 spectral type and short rotation period of $P_{rot}=3.3$~days \citep[][recently revised from 1.4~days; \citealt{Gillon.etal:17}]{Luger.etal:17}.  X-ray observations \citep{Cook.etal:14, Wheatley.etal:17} confirm it has a hot corona with a ratio of X-ray to bolometric luminosity $L_X/L_{bol} = 2$--$4 \, \times10^{-4}$---within the observed scatter around the saturation limit of $L_X/L_{bol}\sim \times10^{-3}$ \citep{Wright.etal:11}. TRAPPIST-1 is then fully expected to have a solar-like wind consistent with the observed spin-down of fully convective M dwarfs \citep[e.g][]{Irwin.etal:11} that could be a destructive agent of planetary atmospheric loss.

In earlier work, we have used a state-of-the-art magnetohydrodynamic (MHD) code to model the stellar winds and magnetospheres of systems around M dwarfs and studied the space environment and atmospheric impact for planets in the "habitable zone" \citep{Cohen.etal:14, Cohen.etal:15, Garraffo.etal:16}.   With its planets even closer in to the star than the case of Proxima and Proxima b considered by \citet{Garraffo.etal:16}, the TRAPPIST-1 system is sufficiently different to warrant further study.
Here,  we use a similar technique to model the space weather conditions of the planets around TRAPPIST-1 and make the further important step of computing the response of the planetary magnetospheric structure to the stellar wind.

%%%%%%%%%%%%%%%%%%%%%%%%%%
%  Model
%%%%%%%%%%%%%%%%%%%%%%%%%%

\section{Magnetohydrodynamic Modeling}
\subsection{Method}
We use the {\it BATS-R-US} MHD code \citep{Vanderholst.etal:14} to model the TRAPPIST-1 stellar system.  The simulation results are obtained using the Alfv\'en Wave Solar Model \citep[AWSoM][]{Vanderholst.etal:14}, which is the Solar Corona (SC) module of the {\it BATS-R-US} MHD code. The model is driven by the photospheric stellar magnetic field data and it solves the set of non-ideal magnetohydrodynamic (MHD) equations on a spherical grid. 

The MHD equations include the conservation of mass, momentum, magnetic flux and energy. In order to provide a hot corona and stellar wind acceleration, two additional equations are solved for the counter-propagating Alfv\'en waves along the two opposite directions of magnetic field lines, where the model accounts for the dissipation of energy as a result of a turbulent cascade. The two equations for the two Alfv\'en waves are coupled to the energy equation via a source term, and to the momentum equation via an additional wave pressure term. 

In the energy equation, the code also accounts for thermodynamic effects, such as radiative cooling and electron heat conduction. Due to the lack of information about the level of MHD turbulence in other stars such as M dwarfs with strong magnetic fields, the model embeds a scaling relation between the stellar field and the total heating via the relation between the observed total magnetic flux and X-ray flux from solar features and stars \citep{Pevtsov.etal:03}. Thus, the parameter that controls the amount of heat flux, $L_{perp}$ in \cite{Vanderholst.etal:14}, scales with the square root of the average magnetic field on the stellar surface. We scale this parameter for the M-dwarf star with respect to the validated value for the solar case. For full details of the model and the references for the theoretical models which are implemented, we refer the reader to \cite{Vanderholst.etal:14}. 
%The scaling of the coronal heating depends on the amount of Poynting flux that is coming from below, and the characteristic length-scale of the perturbation of the field lines. The former takes into account the observed scaling law between the stellar magnetic flux and the stellar x-ray luminosity \citep{Pevtsov.etal:03}, while the latter depends on $\sqrt{B}$, where $B$ is the magnetic field amplitude. For full details of the model and the references for the theoretical models which are implemented, we refer the reader to \cite{Vanderholst.etal:14}. 

%\textcolor{red}{CG: this text instead of the previous paragraph? Due to the lack of information about the level of turbulence in other stars, in particular the with strong magnetic field like M-dwarfs, the model embeds a scaling relation between the stellar field and the total heating via the Pevtsov 2003 relation between the total magnetic flux in the magnetogram and the observed X-ray from the star. Thus, the parameter that controls the amount of heat flux, Lperp in van der holst 2104, scales with the square root of the average magnetic field of the magneto gram. We scale this parameter for the M-dwarf star with respect to the validated value for the solar case.}

The model uses a map of the two-dimensional surface distribution of the stellar radial magnetic field (a ``magnetogram'') as the inner boundary condition. For stars, these magnetograms are typically obtained using high-resolution spectropolarimetry and the Zeeman-Doppler Imaging (ZDI) technique \citep{Semel:80,Donati.Brown:97}.
%,Hussain.etal:01,Piskunov.Kochukhov:02}. 
The initial condition for the three-dimensional magnetic field is obtained by calculating the analytical solution to Laplace's equation assuming that the field is potential, i.e., a static magnetic field with no currents forcing its change. Once the MHD solution starts to evolve, the currents represented by the evolving stellar wind begin to affect the initial, static magnetic field until a steady-state is achieved. By including all the terms above, the stellar input parameters for the mass, radius, and rotation period, as well as the photospheric field data, the model provides a self-consistent, steady-state solution for the hot corona and stellar wind from the stellar chromosphere out to the extent of the adopted model grid.
%to a distance of about $30\;R_\star$.  
The models presented here employed adaptive mesh refinement with a maximum resolution of $0.05 R_\star$.

\citet{Evans.etal:08} compared different models for coronal heating and wind acceleration including empirical, semi-empirical, and Alfv\'en wave heating. They concluded that the only type of models that provide good agreement with coronal signatures are those with Alfven wave heating, as employed by the AWSoM model.
This numerical approach is currently used for space weather forecast in the solar system, and has been validated against observations in several works \citep[e.g.,][]{Meng.etal:15,Alvarado-Gomez.etal:16a,Alvarado-Gomez.etal:16b}. Furthermore, AWSoM has been tested and validated by extreme ultraviolet observations of Sun and its immediate vicinity \citep{Vanderholst.etal:14}, which is the regime we wish to explore in this work for the close-in planets in TRAPPIST-1.

In order to study the interaction between the extreme stellar wind and the upper atmosphere of one of the TRAPPIST-1 planets, we use the Global Magnetosphere module of {\it BATS-R-US}. This module is driven by the upstream stellar wind conditions as extracted from the AWSoM model. 

\subsection{Calculations}

\begin{figure*}[h]
\center
	%\includegraphics[trim = 0.5in 4in
	%  0.8in 0.8in,clip, width =  0.5\textwidth]{Star_zoom.png}
\includegraphics[trim = 0.in 0in
0.in 0.in,clip, width =  0.8\textwidth]{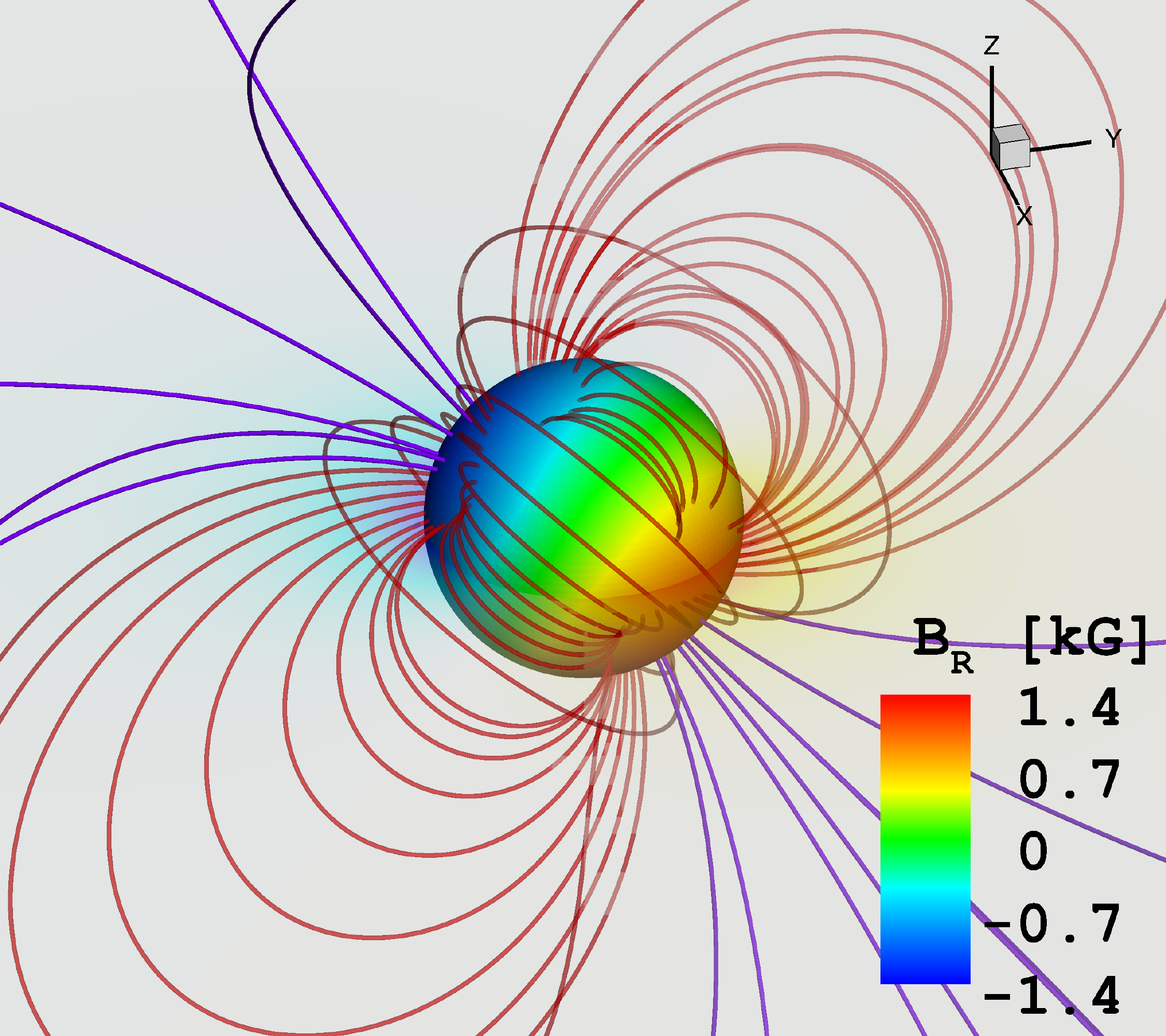}
	%  \includegraphics[trim = 0.in 0in
	%  0.in 0.in,clip, width =  0.8\textwidth]{Fig1_V3.pdf} 
	%\includegraphics[trim = 0.5in 0.in
	 % 0.3in 0.in,clip, width =  0.95\textwidth]{trappist_B1_Br.png}
\caption{Magnetogram for GJ 3622 with $600$~G average field strength. Closed field lines are colored in red and open field lines in purple.}
\label{fig:magnetogram}
\end{figure*}

Unfortunately, TRAPPIST-1 is too faint ($M_{\rm V} = 18.4$) to obtain ZDI maps with current instrumentation. However, the average magnetic field strength was measured using Zeeman Broadening to be $600$~G \citep{Reiners.Basri:10}. 
Empirically, the distribution of surface magnetic field at a given spectral type is found to depend mainly on rotation rate \citep{Vidotto.etal:14b,Reville.etal:15a,Garraffo.etal:15}. To model the system we can then use as a proxy the ZDI magnetogram of a star with parameters most similar to those of TRAPPIST-1 available. We adopt the magnetogram of GJ~3622 \citep{Morin.etal:10}, an M6.5 dwarf with a rotation period of 1.5~days, shown in Figure~\ref{fig:magnetogram}.  In addition, we have performed  simulations for the recently revised rotation period of 3.3~days \citep{Luger.etal:17} and find results are unaltered. This is expected since the magnetic field strength estimation is unchanged and other effects of fast rotation, like magnetic field wrapping, only becomes important at shorter rotation periods of less than a day. We use the observed relative orientation of the magnetic fields with respect to the rotation axis. The range of mean surface field strengths allowed by the \citet{Reiners.Basri:10} measurement is about 200-800~G. In order to understand the influence of the mean surface magnetic field strength on the results, we probe two magnetic field scalings of the magnetogram, one to match the $\sim 600~G$ surface field measurement, and one with half of that value.

%Stellar magnetic fields, due to their high energy density, play a major role in heating the stellar corona similarly to the solar regime and then the hot corona that drives the stellar wind \citep{Owens.Forsyth:13}.  

%\cite{Morin.etal:10} argue that the magnetic field determination is not uniquely determined by mass and period. This might be true for the magnetic field strength determination via ZDI, which usually results in much lower fields than expected and measured with other techniques, due to field cancellation. However, we have ran the spherical harmonics decomposition for the stars in that work and we find that the complexity of the field (the dominant multipole order) is a well behaved function of their rotation period. Therefore, using the ZDI map of a star of the same rotation period and similar spectral type, and scaling the field strength to the Zeeman Broadening expected, consistent with X-ray observations, should provide a reliable proxy for TRAPPIST-1. 

As a test, we also computed models for the magnetogram of the very late dwarf VB~10 \cite{Morin.etal:10} using the TRAPPIST-1 rotation period.  While VB~10 has the same M8 spectral type as TRAPPIST-1, its own rotation period is currently uncertain, with values of 0.52 and 0.69 days favored. The magnetogram was reconstructed for the latter but
is consequently subject to considerable uncertainty. We found wind conditions at the TRAPPIST-1 planetary orbits to be quite similar to those of our GJ~3622  proxy model. The reason is that the dominant factor responsible for the extreme space weather environment in these kind of systems is the high plasma densities and pressures the close-in planets reside in, which in turn depend on the stellar magnetic field strength.  For VB~10, this is similar to that of our proxy. We do not discuss the VB~10 results further, and instead concentrate on our TRAPPIST-1 proxy simulations. 

The MHD model of the stellar corona, wind, and magnetic field of TRAPPIST-1 was driven using its measured mass, radius and rotation period, $M = 0.08~M_{\odot}$, $R = 0.114~R_{\odot}$, and $P_{rot} = 1.4$~days, respectively.
From the resulting three-dimensional wind structure, illustrated in  Figure~\ref{fig:3d}, we extracted the wind pressure values over all the planetary orbits. The semi-major axes are known and range from 0.011~AU to 0.063~AU, all the eccentricities are constrained to be smaller than 0.085, and the inclinations of the orbits with respect to the observer's line of sight are nearly 90 degrees.  We assume that this very nearly coplanar system of planets has an orbital axis aligned with the star's rotation axis. Therefore we model the seven circular orbits with their observed planetary parameters on the equatorial plane.

All the planets detected around TRAPPIST-1 have Earth-like masses, and we examine planetary magnetic fields with equatorial strengths of 0.1~G and  0.5~G that bracket the present day terrestrial value of 0.3~G. 

We are confident that we have produced the most realistic wind model one can currently make for TRAPPIST-1. Mass loss and angular momentum loss rates can be used to calculate spin-down timescales.  We find mass loss rates of $\sim 3\times 10^{-14}~M_\odot$~yr$^{-1}$ and angular momentum loss rates of $ \sim 6 \times 10^{30}$~erg, which are expected for a rapidly rotating M dwarf and consistent with the currently quite uncertain picture of their rotational evolution timescales \citep[see, e.g.,][]{Basri.Marcy:95,West.etal:08,Irwin.etal:11}.

To assess the magnetospheric response of the TRAPPIST-1 planets to the stellar wind using the Global Magnetosphere module, we examine the case of Trappist-1 f.  This is the central planet of the three potentially habitable planets (e, f, and g). We extract the wind conditions at one sub-Alfv\'enic point and one super-Alfv\'enic point along the orbit. Once the upstream conditions are set at the outer boundary of the simulation domain, a steady-state solution for the magnetosphere is obtained. The inner boundary is set at $r=2R_{\oplus}$, and the boundary conditions we assume here are the same as those used in Earth magnetosphere simulations. A model for the Ionospheric Electrodynamics is also used to better set the velocity at the boundary as described in \cite{Cohen.etal:14}.
 
% corresponds to a spin-down timescale of  $\sim 10$~Gyr, consistent with observations and expectations for late M dwarfs \citep{Basri.Marcy:95}.

%{\color{blue} JJD: put in ref for validation/calibration of model on solar wind} \textcolor{red}{JDAG: Maybe we can put the corona and wind papers from last year here as one of these validation references (http://adsabs.harvard.edu/abs/2016A\%26A...588A..28A; http://adsabs.harvard.edu/abs/2016A\%26A...594A..95A)? I haven't seen that many papers validating the wind results using AWSoM (shameless self-citation). Besides, we are the cover of the 594 volume of A\&A (that has to count!) ;) }

%%%%%%%%%%%%%%%%%%%%%%%%%%
%  Results & discussion
%%%%%%%%%%%%%%%%%%%%%%%%%%

\section{Results and Discussion}

\begin{figure*}[h]
\center
\includegraphics[trim = 1in 3.5in
  0.3in .8in,clip, width = 0.48\textwidth]{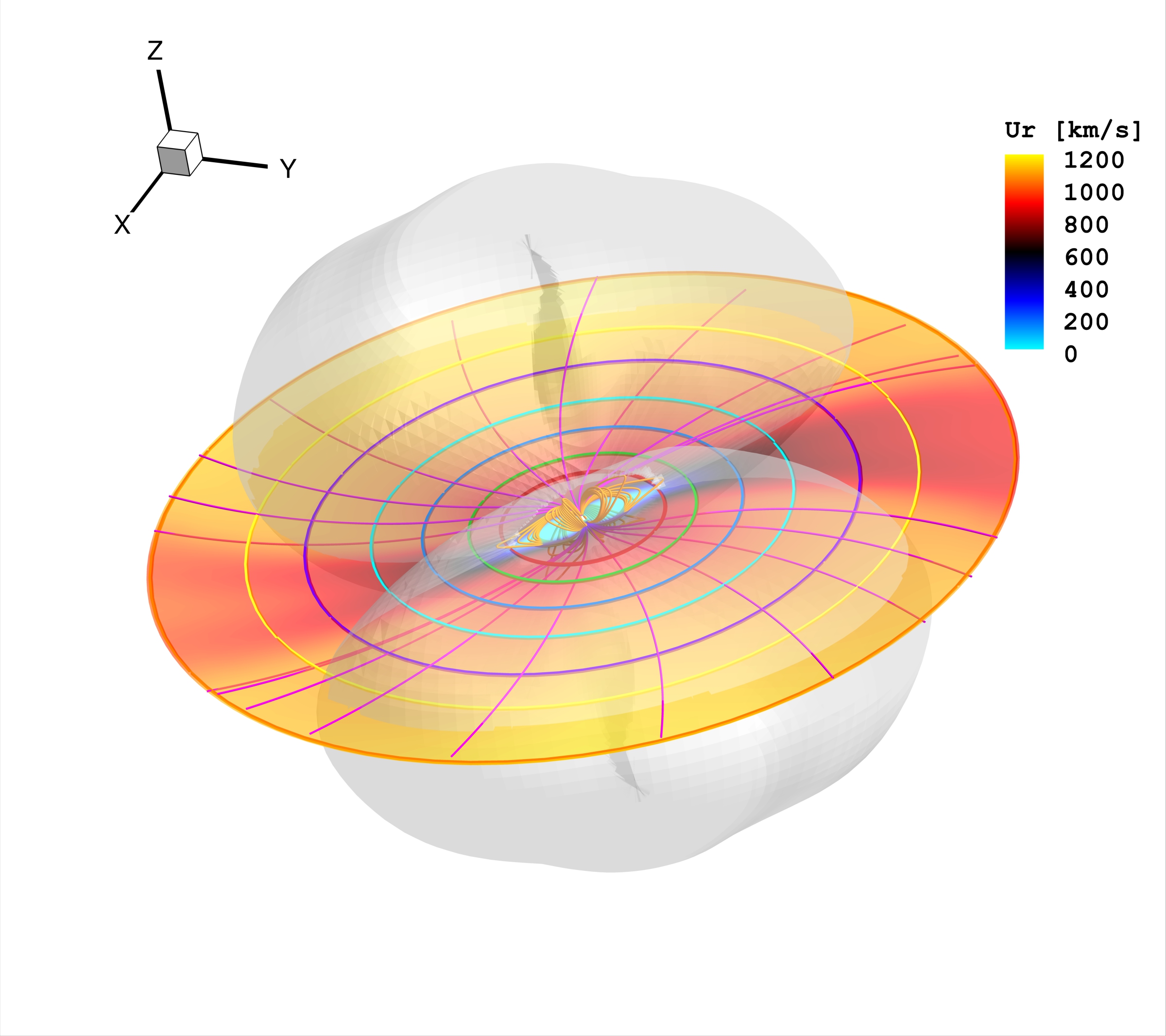}\includegraphics[trim = 1in 3.5in
  0.3in .8in,clip, width = 0.48\textwidth]{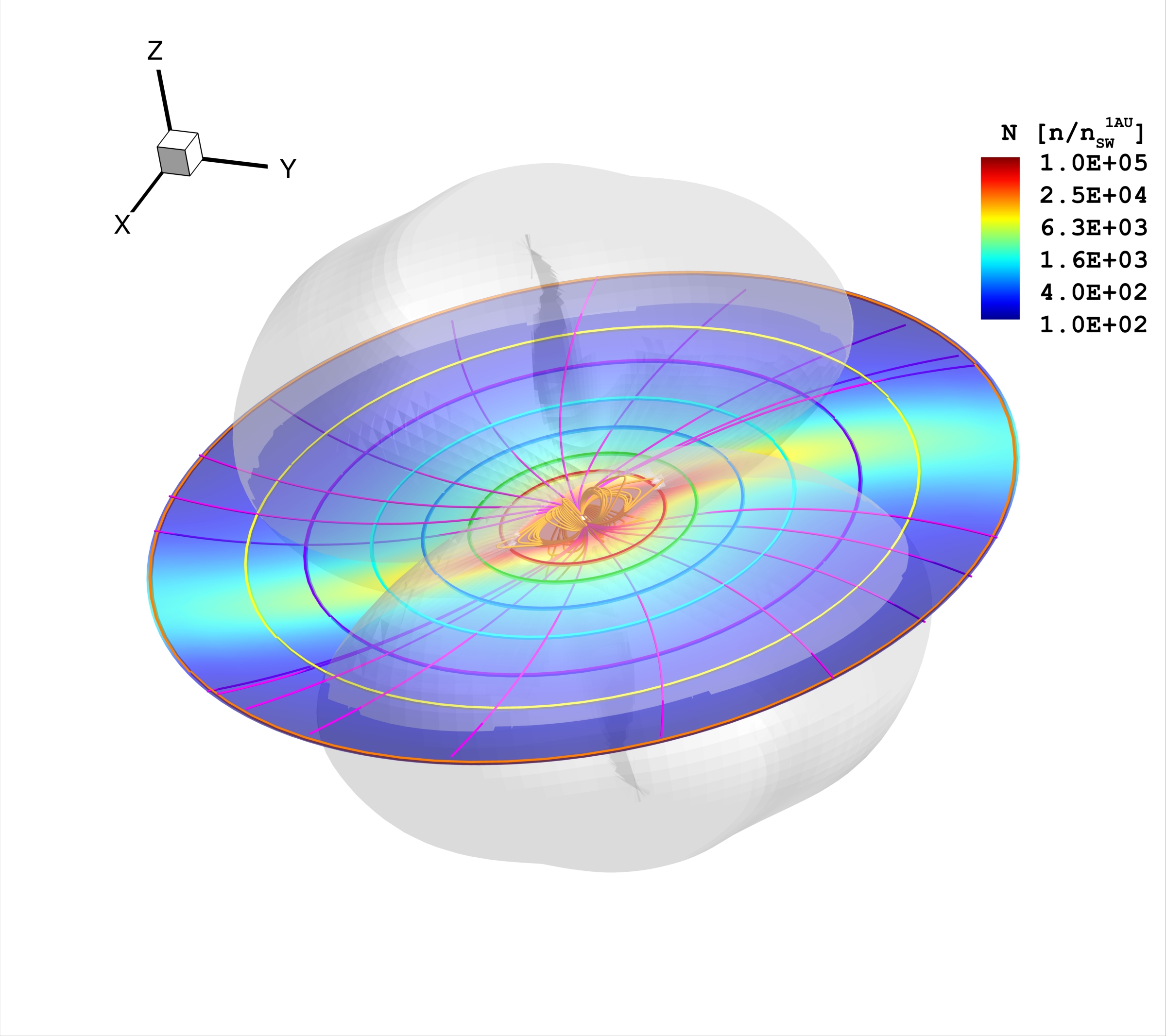}
   \llap{\raisebox{0.25cm}{%  move next graphics to top right corner
      \includegraphics[height=2.4cm]{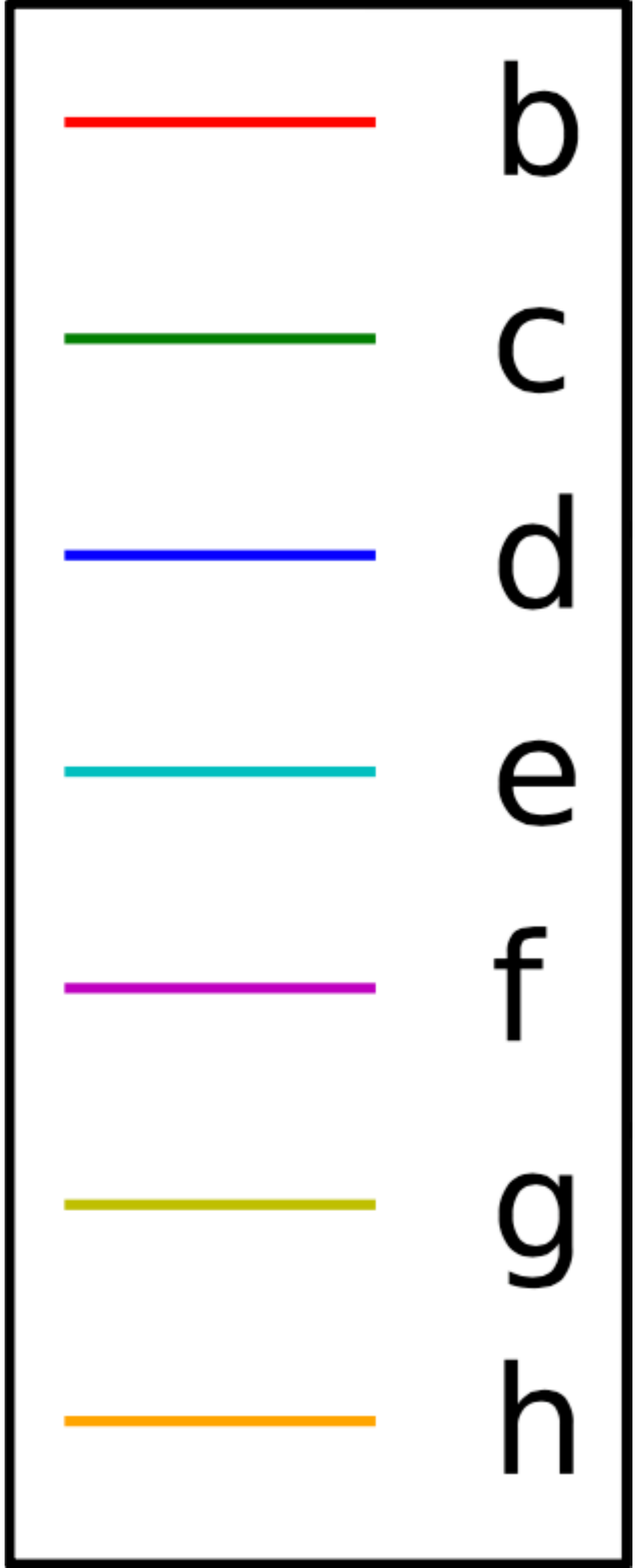}
      }}
\includegraphics[trim = 1in 3.5in
  0.3in .8in,clip, width = 0.48\textwidth]{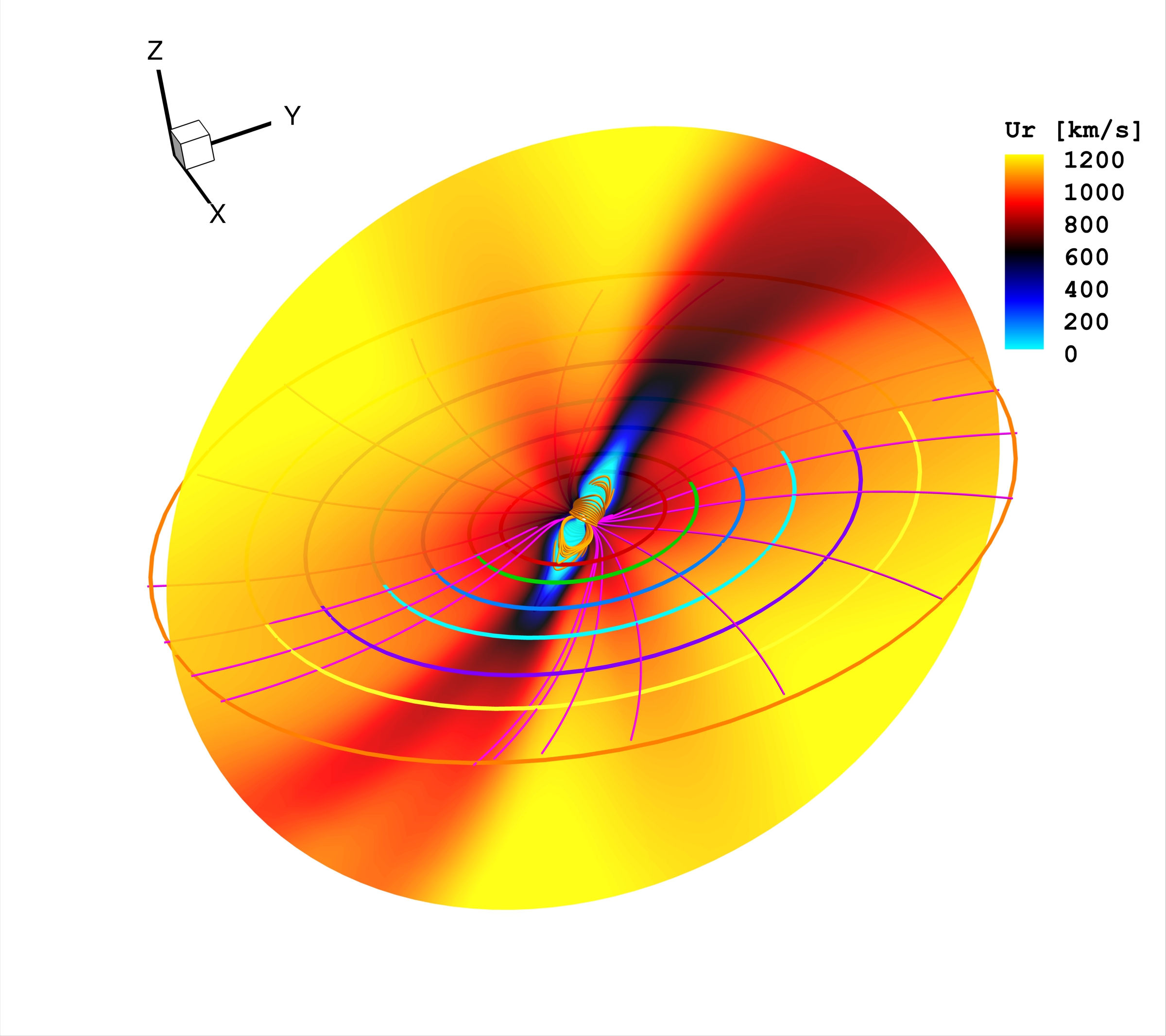}\includegraphics[trim = 1in 3.5in
  0.3in .8in,clip, width = 0.48\textwidth]{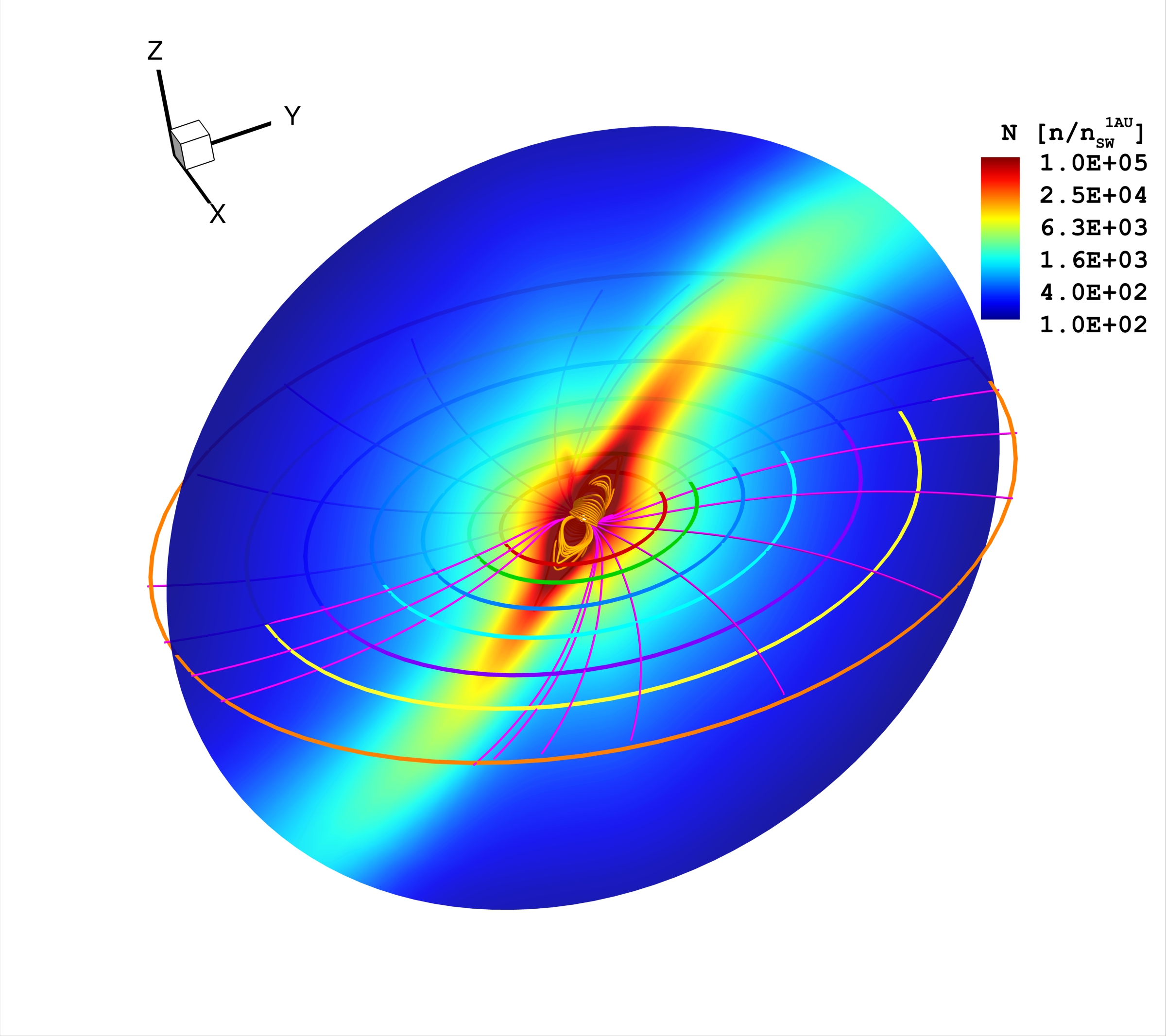}
   \llap{\raisebox{0.25cm}{%  move next graphics to top right corner
      \includegraphics[height=2.4cm]{OrbitsLabel}
      }}
\includegraphics[trim = 1in 3.5in
  0.3in .8in,clip, width = 0.48\textwidth]{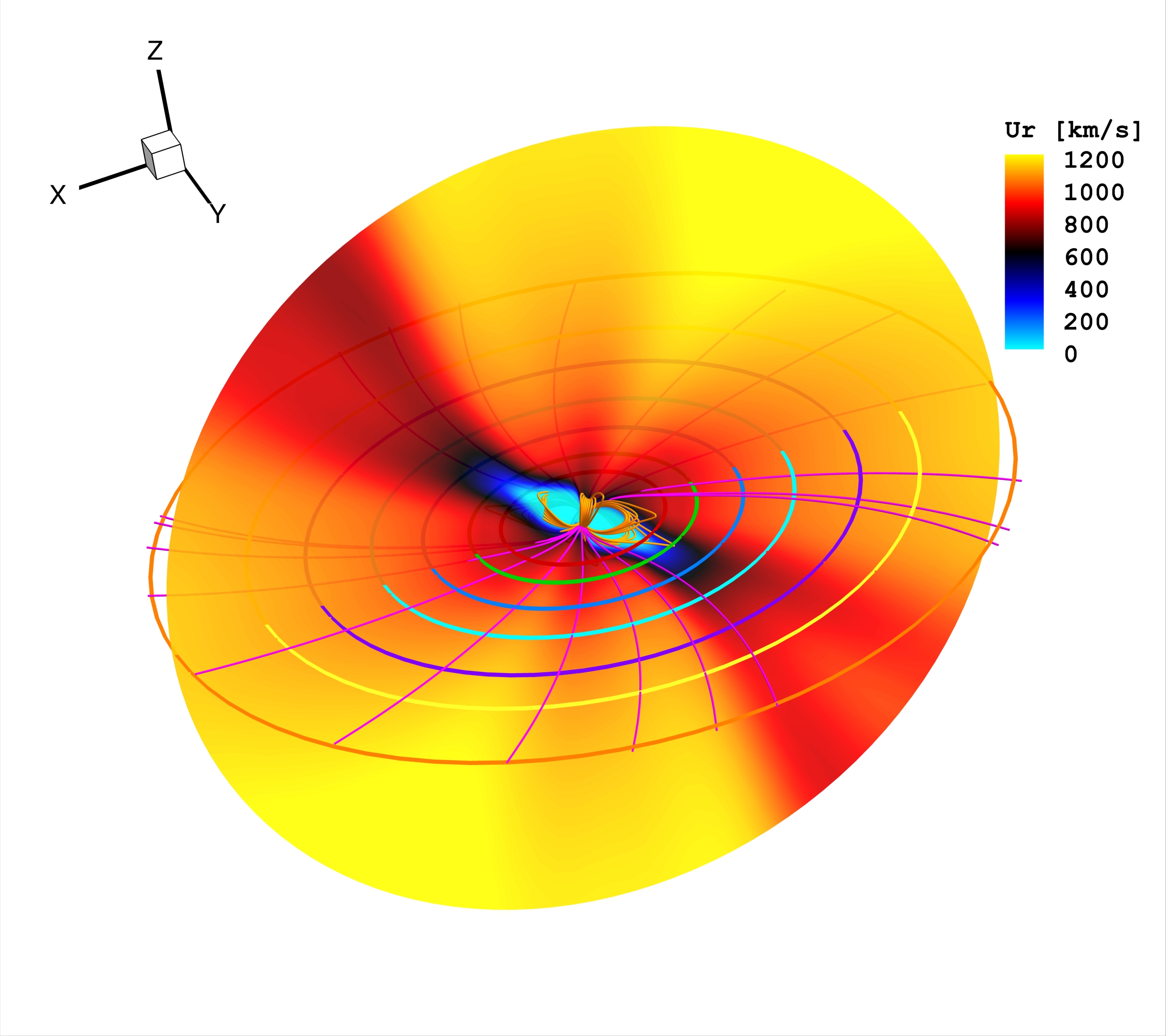}\includegraphics[trim = 1in 3.5in
  0.3in .8in,clip, width = 0.48\textwidth]{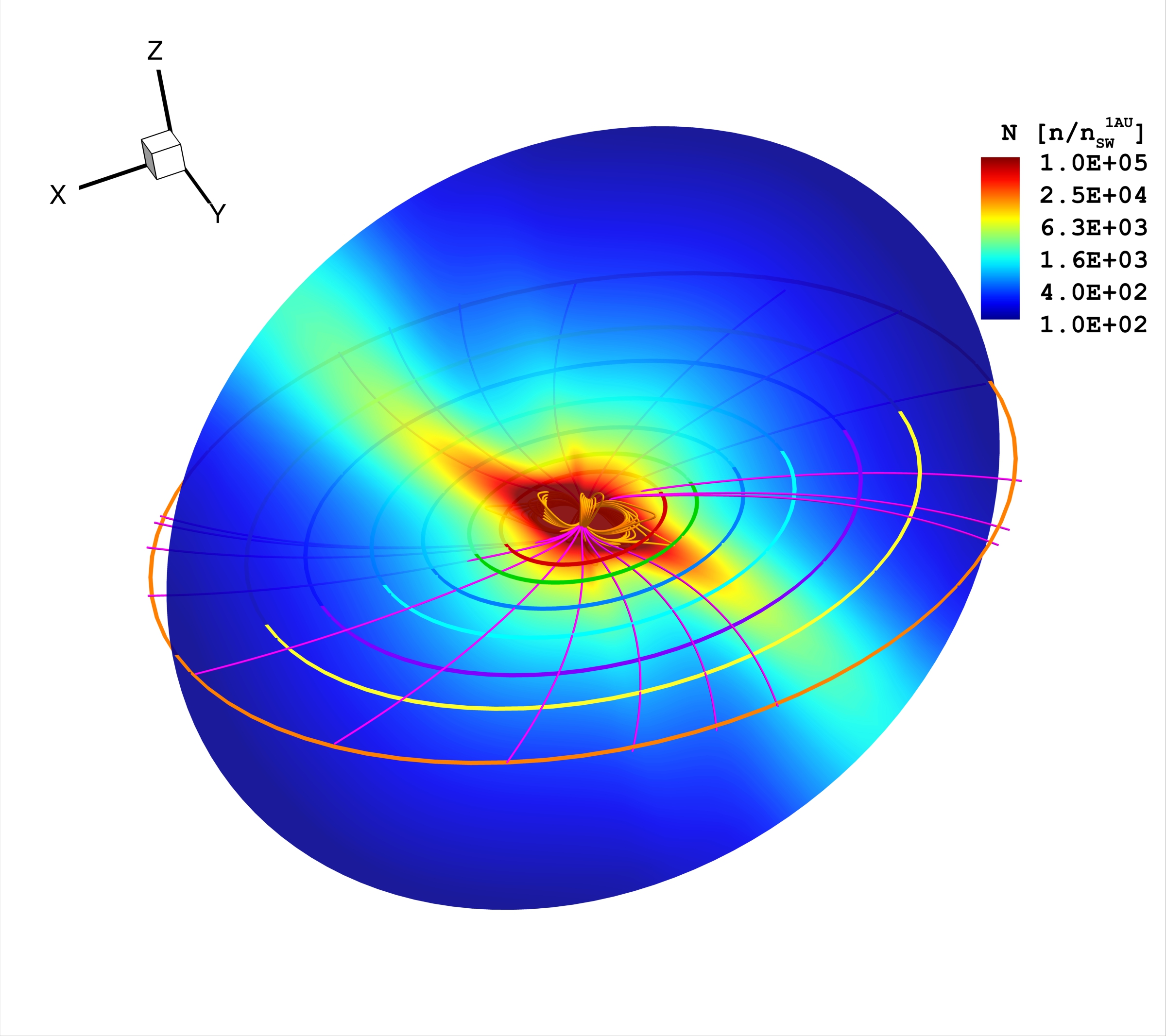}
   \llap{\raisebox{0.25cm}{%  move next graphics to top right corner
      \includegraphics[height=2.4cm]{OrbitsLabel}
      }}
\caption{Three-dimensional stellar magnetosphere and wind for TRAPPIST-1 simulated using a magnetogram for the proxy star GJ 3622 with an average field strength of 600~G. The orbital plane (top row) and two meridional cuts ($x=0$, middle; $y=0$, bottom two rows) are presented. The color bar denotes the wind speed (left), and density normalized to solar wind values at 1~AU (right). The gray shaded surface in the top panels corresponds to the Alfv\'en surface. Selected closed (orange) and open (purple) magnetic field lines are shown. All the known orbits are plotted.}
\label{fig:3d}
\end{figure*}

Figure~\ref{fig:3d} illustrates the three-dimensional wind speed and density for TRAPPIST-1 from our simulations corresponding to the $600$~G observed average magnetic field strength. 
The seven known planetary orbits are plotted, together with the three-dimensional Alfv\'en surface. 
Wind speeds reach close to 1400~km~s$^{-1}$ 
%it is slightly lower for the stronger ($\sim 1250~km\, s^{-1}$)}  
and are only slightly higher than the 800--900~km~s$^{-1}$ typical of the fast solar wind \citep{McComas.etal:07}. In contrast, the densities of the plasma these planets go through reach $10^4$-$10^5$ times the solar wind density at 1~AU. %\textbf{This shows that the density dominates the difference in magnetic ??}
In addition, the planets lie much closer to the Alfv\'en surface than in the solar system. The solar wind Alfv\'enic critical point generally lies between 6 and 20$R_\odot\approx0.03-0.1$AU \citep{DeForest.etal:14} ---well within the orbit of Mercury.  Instead, all but the outermost TRAPPPIST-1 planets spend a large fraction of their orbits in the sub-Alfv\'enic regime, crossing the Alfv\'en surface four times over their short orbital periods ($< 13$~days). 

\begin{figure*}[h]
\includegraphics[trim = 2in 2.5in
  3.3in 2.5in,clip, width = 0.48\textwidth]{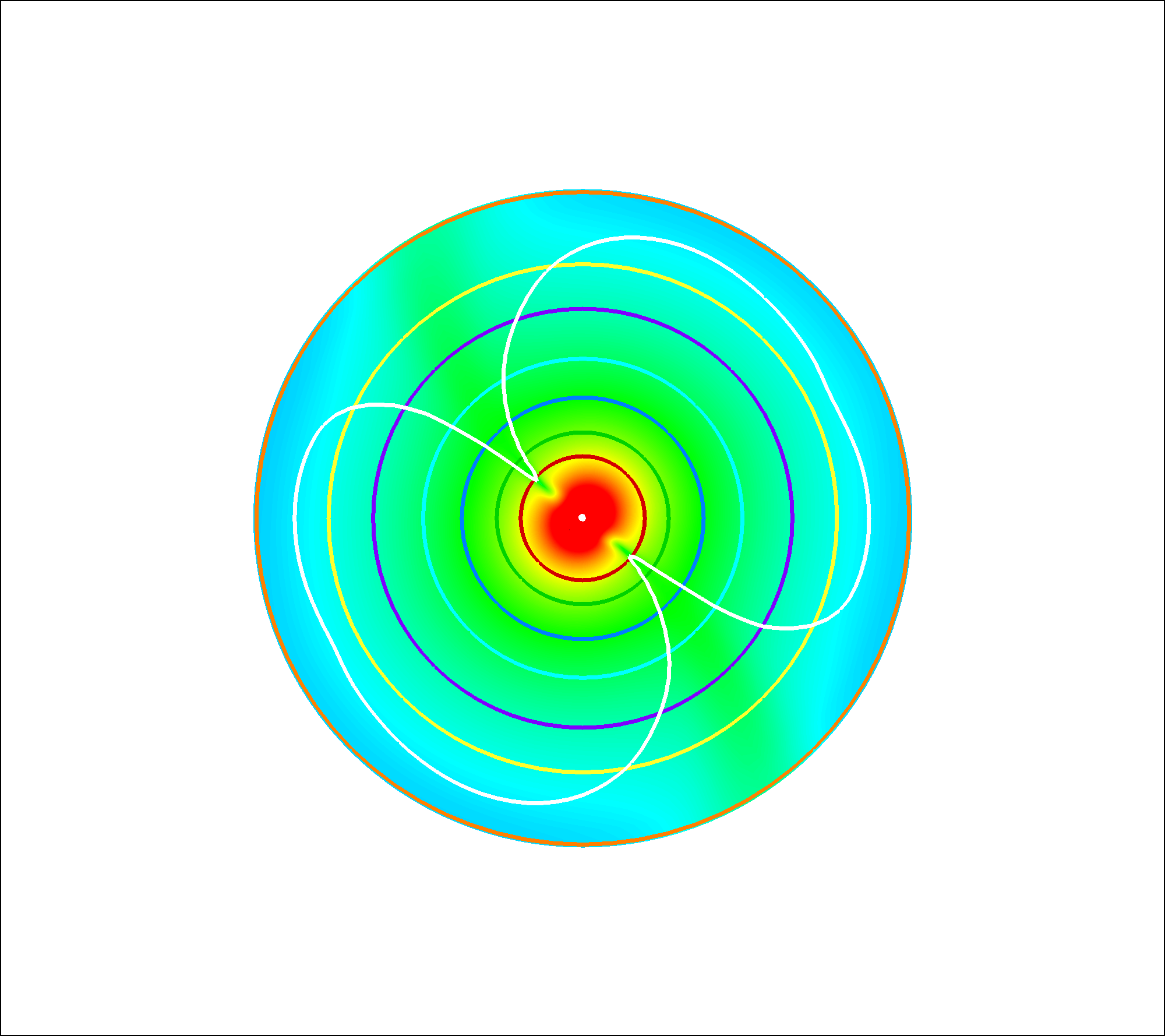}
 \includegraphics[trim = 5in 2.5in
  .3in 2.5in,clip, width = 0.48\textwidth]{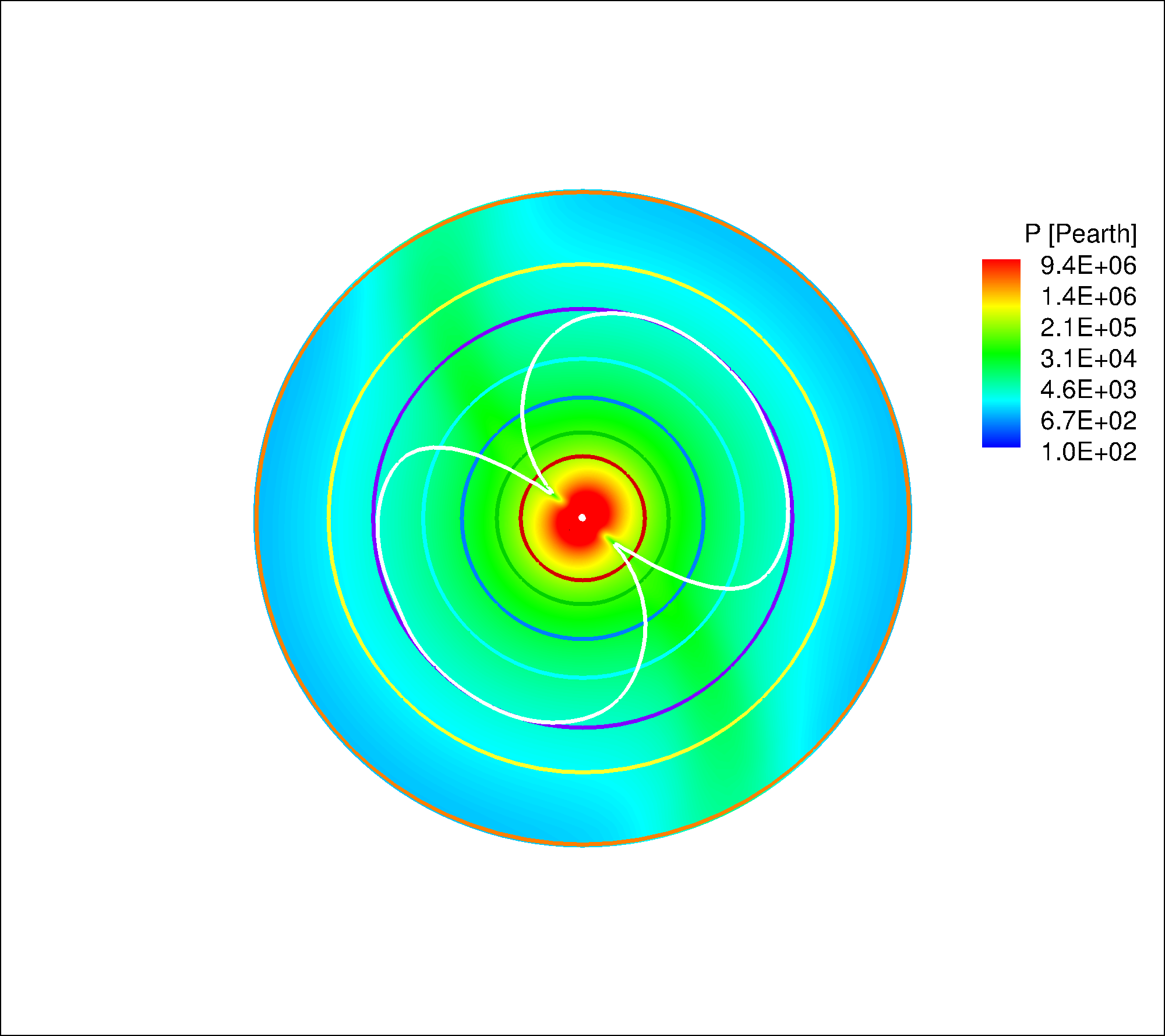}
 \llap{\raisebox{1cm}{%  move next graphics to top right corner
      \includegraphics[trim = 0.in 0in
  0in 0in,clip, width = 0.05\textwidth]{OrbitsLabel}%
    }}
\caption{Total ambient pressure normalized to the solar wind pressure at 1 AU for all the detected orbits of TRAPPIST-1 for a mean stellar magnetic field of $600$~G (left) and $300$~G (right).  The Alfv\'en surface is shown in white. }
  \label{fig:2d}
\end{figure*}

In Figure~\ref{fig:2d} we show the total ambient pressures (magnetic plus dynamic, neglecting thermal and orbital terms that are at least an order of magnitude smaller), $P_{\rm tot} = B^2/(4 \pi)+ \rho U^{2}$, where $B$ is the magnetic field strength, $\rho$ is the plasma mass density and $U$ is the wind speed, normalized to that of the solar wind at Earth for each magnetic field scaling. We also show the Alfv\'en surface intersection with the orbital plane and the seven detected orbits. The total pressure at these close-in orbits ranges from 3 to 6 orders of magnitude higher than the solar wind pressure at 1~AU. 
%\textbf{This is the result of the much higher densities the planets are exposed to rather than the wind speeds' increase, as seen from Figure~\ref{fig:3d}}. 
In addition, even in the presence of wind speeds comparable to the solar wind one, due to the extremely high densities of the plasma around these close-in planets shown in Figure~\ref{fig:3d}, the ambient dynamic wind pressure they are exposed to is 3-4 orders of magnitude larger than the solar wind pressure at Earth.  

\begin{figure*}[h]
\center
%\vspace{-0.2in}  
\includegraphics[trim = .1in 0.in
  0.2in 0.3in, clip, width =  0.49\textwidth]{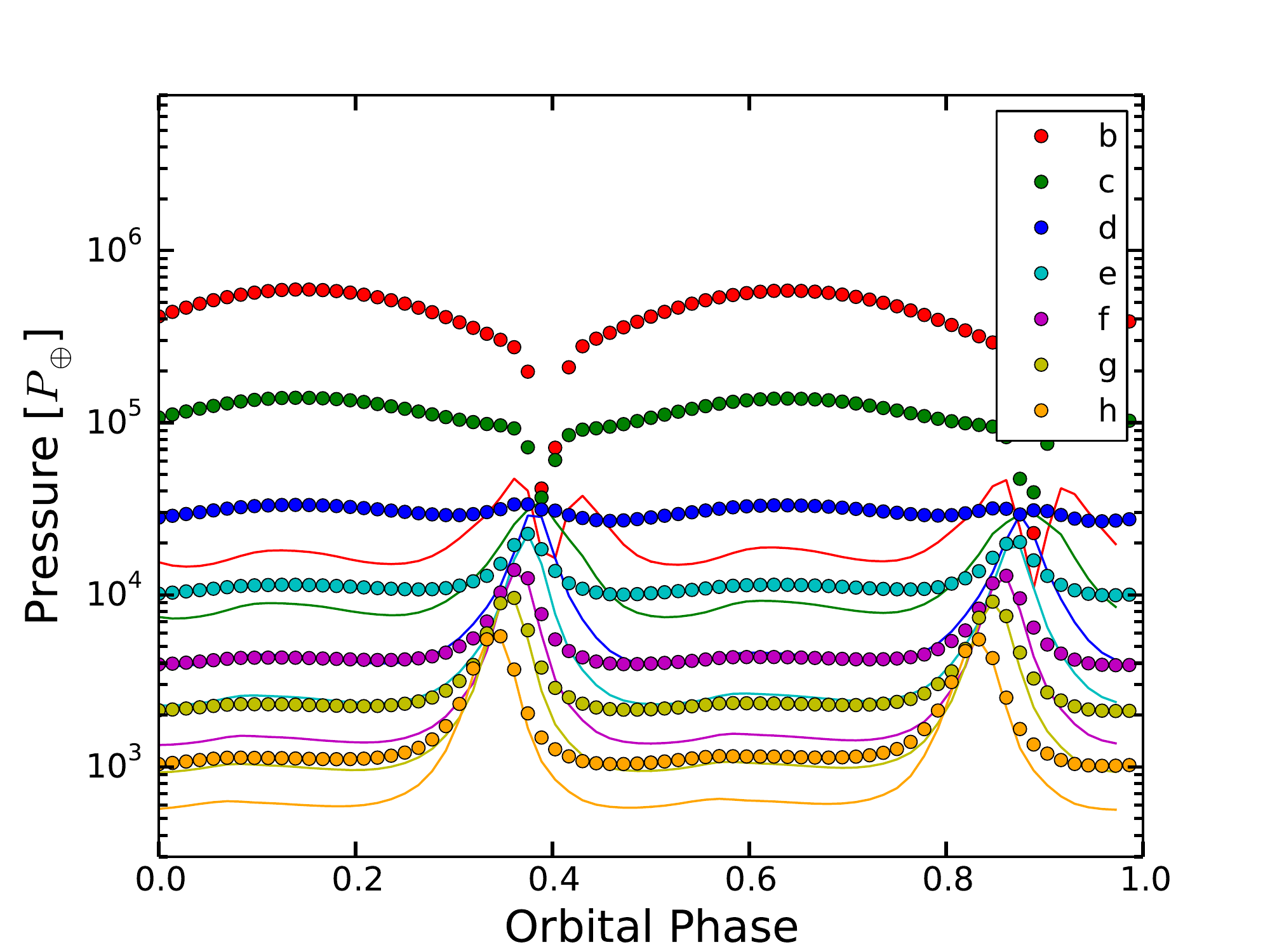}
  \includegraphics[trim = .1in 0.in
  0.2in 0.3in, clip, width =  0.49\textwidth]
  {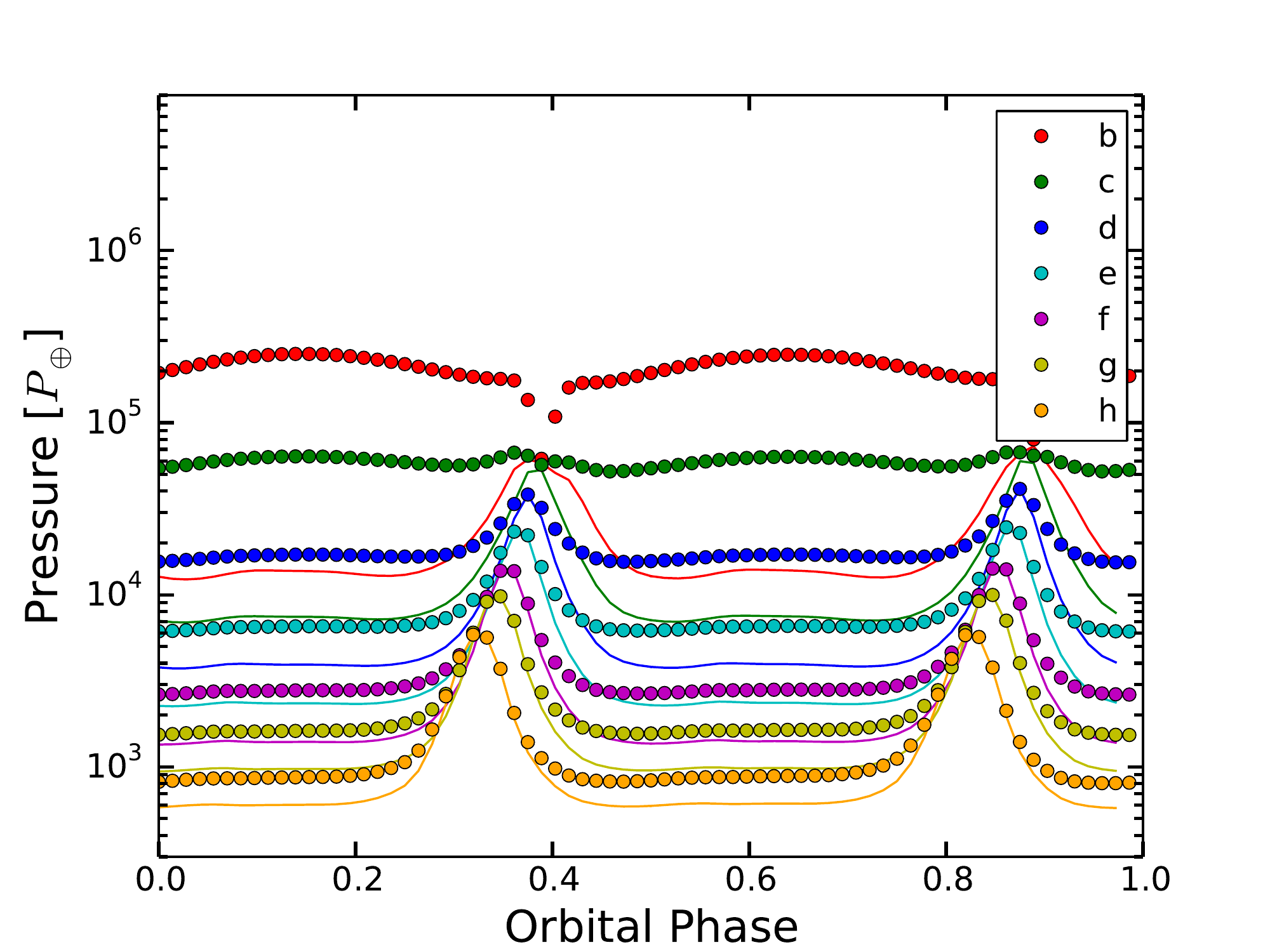}\\
  \includegraphics[trim = .1in 0.in
  0.2in 0.3in, clip, width =  0.49\textwidth]{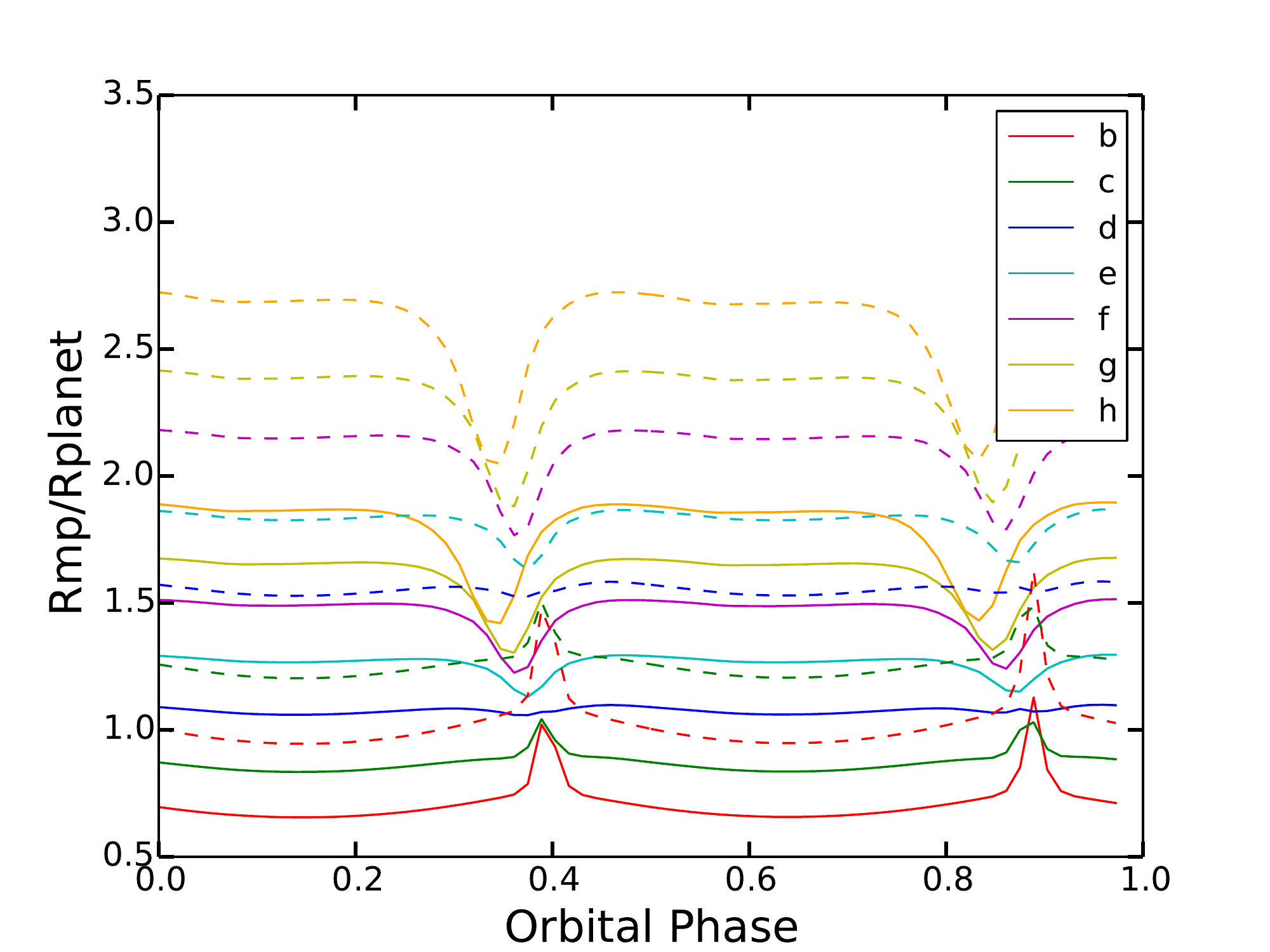}
 \includegraphics[trim = .1in 0.in
  0.2in 0.3in, clip, width =  0.49\textwidth]{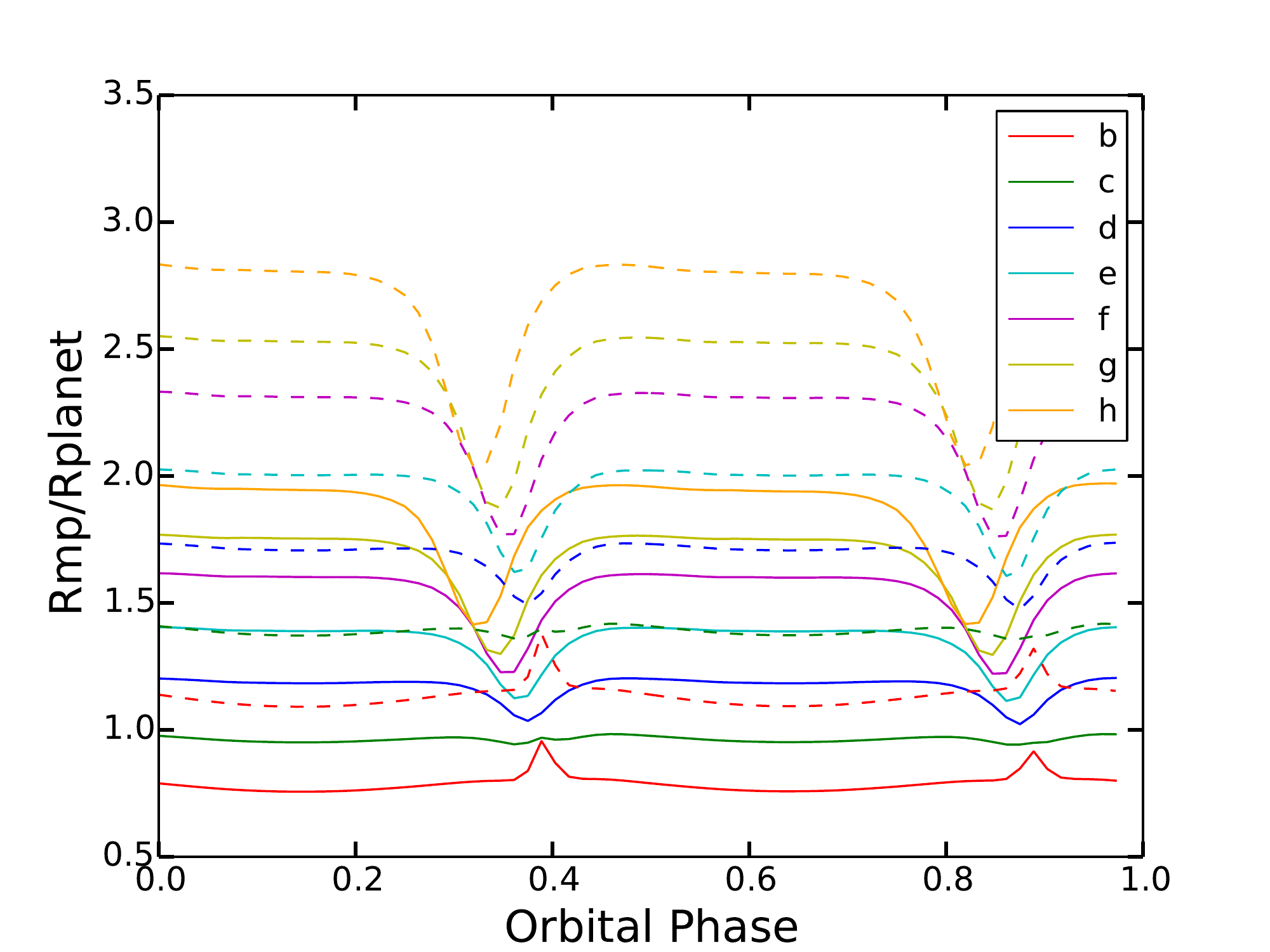}\\
\caption{Top panel: Total pressure (filled circles) and dynamic pressure (lines) normalized to the solar wind at $1$~AU for all the planetary orbits assuming the measured stellar magnetic field strength of $\sim 600~G$ (left) and half of that ($\sim 300~G$, right).  
Bottom panel: Magnetosphere standoff distance normalized to the planet's radius for all the planetary orbits for the same $\sim 600~G$ (left) and $\sim 300~G$ (right) stellar magnetic fields. Dashed and solid lines correspond to a $0.5$~G and $0.1$~G planetary magnetic field respectively.}
\label{fig:orbits}
\end{figure*}

Both the total and dynamic pressure over the seven orbits for each of the two explored magnetic field scalings are illustrated in more detail in the top panel of Figure~\ref{fig:orbits}.  For the closer-in planets, the magnetic pressure dominates over the dynamic pressure, while the converse tends to be true for the very outer planets.
%That is the reason for the difference in the shapes of the curves over the orbital phase.
%We find that for the two studied cases, the stellar dynamic wind pressure over all orbits reaches values of 4--5 orders of magnitude the solar wind pressure at Earth. 
In addition to the extreme pressure, we see that most orbits go through magnetic and dynamic pressure variations of up to an order of magnitude crossing the current sheet in the vicinity of the magnetic equatorial plane.
%for the outermost planet. 
For planet b, with an orbital period of only 1.51 days \citep{Gillon.etal:17}, this happens on a timescale of only 3--4 hours. The most stable total pressure is along the orbits of planet d for the 600~G stellar field and planet c for the 300~G field, for which the large increase in dynamic pressure when crossing the current sheet compensates for the dip in magnetic pressure.

One diagnostic of the effect of wind conditions on a magnetized planet is the magnetopause location, whose 
%(assuming that the planets are magnetized with an Earth-like field of $\sim 0.5~G$). The magnetopause 
standoff distance from the planet can be approximated by assuming pressure balance between the planetary magnetic pressure and the wind total pressure \citep[e.g.,][]{Schield:69,Gombosi:04} $$R_{\rm mp}/R_{\rm planet}=[B_p^2/(4\pi P_{\rm tot})]^{1/6},$$ where $R_{mp}$ is the radius of the magnetopause, $R_{\rm planet}$ is the radius of the planet, $B_{\rm p}$ refers to the planetary equatorial magnetic field strength, and $P_{\rm tot}$ is the ram pressure of the stellar wind combined with the stellar magnetic field pressure. The magnetopause distance as a function of orbital phase for the seven different orbits considered and for the two stellar magnetic field scalings is illustrated in Figure~\ref{fig:orbits} (bottom panel).  If these planets have magnetospheres, they would be much smaller than that of the Earth, which has a standoff distance of $\sim$10$R_{\rm \oplus}$ \citep{Pulkkinen.etal:07}. However, according to our simulations, most of the TRAPPIST-1 planets reside in the sub-Alfv\'enic regime for large fractions of their orbital period.  For this reason, we go a step further and simulate directly the wind-exoplanet interaction using a global magnetosphere model.

%The magnetosphere sizes are much smaller than that of the Earth, which has a standoff distance of $\sim$10$R_{\rm \oplus}$\cite{Pulkkinen.etal:07}. 
%Even for the strongest planetary magnetic field considered ($\sim 0.5~G$) the magnetosphere sizes are smaller than 3 planetary radii for all cases and orbits. In addition, most magnetospheres go through fast size variations, except for planet d for the for the 600~G stellar field and planet c for the 300~G field, as expected from the stable pressure they are exposed to over their orbits. 

%Such intense pressures on the planetary orbits can result in strong atmospheric stripping, small magnetospheres in poor atmospheric protection, and fast variations in their magnetospheric sizes are expected to drive strong currents that can lead to atmospheric heating and evaporation. 

\begin{figure*}[h]
\center
\includegraphics[width =  0.8\textwidth]{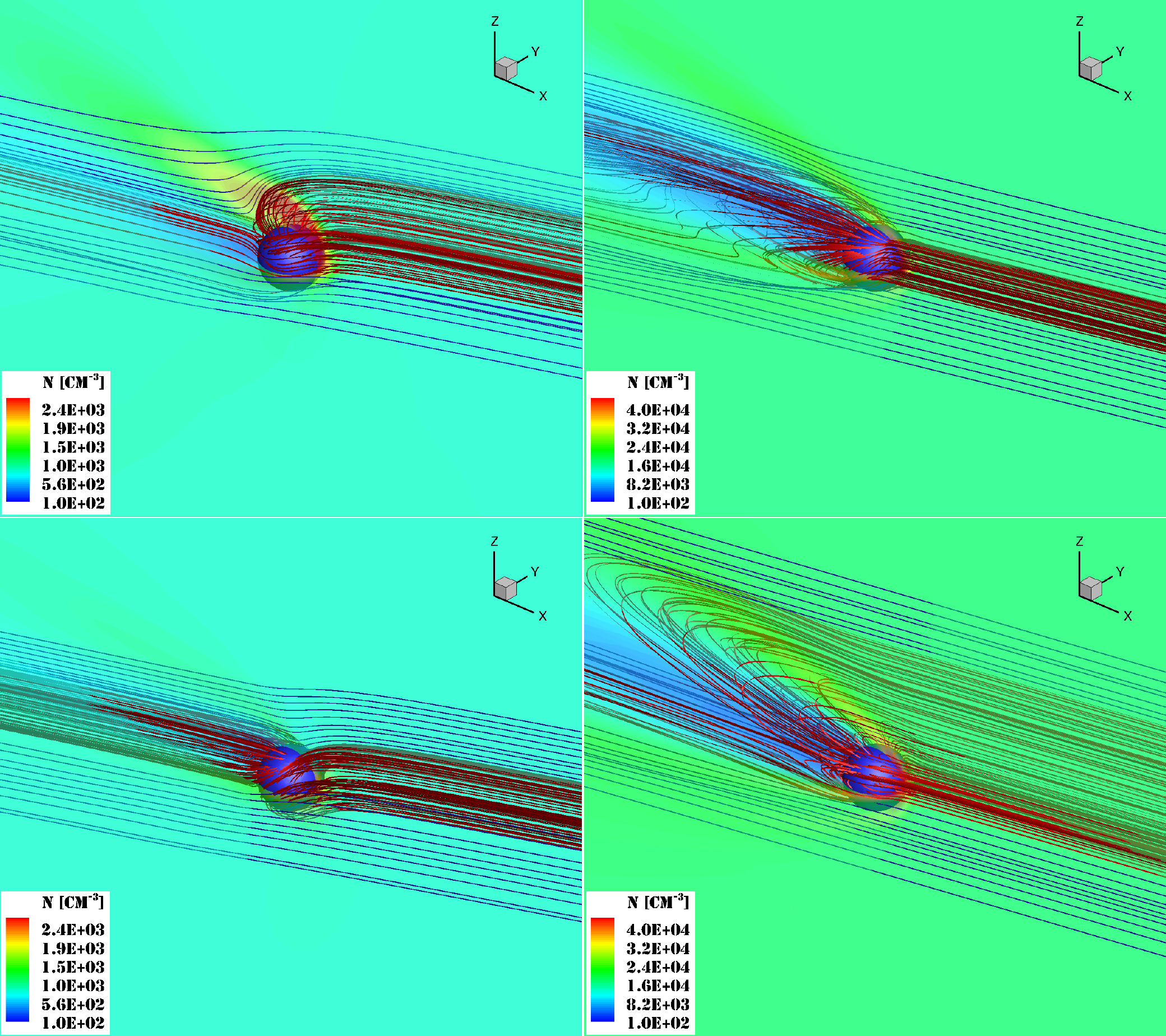}
\caption{Magnetospheric structure of TRAPPIST-1 f assuming a planetary magnetic field the same as the that of the Earth (top panels), and an Earth field with an opposite sign (bottom panels). The star is located on the right along the $x$ axis and the sphere represents the planet. The solutions are for sub-Alfv\'enic (left panels) and super-Alfv\'enic (right panels) stellar wind conditions as extracted from the stellar wind model. Field lines that rae connected to the planet are shown in red while nearby stellar field lines are shown in blue. These results are for a stellar field of 600G. The magnetospheric solutions for the 300G field were very similar.}
\label{fig:GM}
\end{figure*}

Figure~\ref{fig:GM} shows the magnetospheric structure of TRAPPIST-1 f as calculated using the Global Magnetosphere module of {\it BATS-R-US} . Trappist-1 f provides a representative case of the three potentially habitable planets e, f, and g, and from Figure~\ref{fig:orbits} it can be seen that the stellar wind conditions around the three are quite similar. Here, we make a specific assumption that the planet is magnetized and the planetary field is similar to that of the Earth as there is no available data to constrain the planetary field.

%In both cases of the stellar wind drivers (i.e. different stellar magnetic field scalings), it can be seen that 
Planet f resides in a region which is dominated by strong radial components of both the stellar wind velocity and magnetic field. The extreme wind pressure (dynamic and magnetic in the case of the sub-Alfv\'enic regions) opens the planetary field all the way to the planetary surface, creating what is essentially a very large polar cup (open field region) that extends over most of the planet. This is a new regime not experienced in solar system planets: there is no magnetopause at which the planetary field pressure balances the wind pressure. Instead, stellar wind particles can constantly precipitate directly down open field onto the atmosphere.      
The concept of atmospheric protection by a planetary magnetic field does not hold here and is likely not to hold in the conventional sense for the TRAPPIST-1 planets. The TRAPPIST-1 system represents a new challenge to atmospheric evolution and survival on close-in planets around very low mass stars.

\acknowledgments

CG thanks Rakesh K. Yadav for useful comments and discussions. CG was supported by SI Grand Challenges grant ``Lessons from Mars: Are Habitable Atmospheres on Planets around M Dwarfs Viable?''.  JJD was supported by NASA contract NAS8-03060 to the {\it Chandra X-ray Center}.  OC and SPM are supported by NASA Astrobiology Institute grant NNX15AE05G. JDAG was supported by {\it Chandra} grants AR4-15000X and GO5-16021X.
This work was carried out using the SWMF/BATSRUS tools developed at The University of Michigan Center for Space Environment Modeling (CSEM) and made available through the NASA Community Coordinated Modeling Center (CCMC). Simulations were performed on NASA's PLEIADES cluster under award SMD-16-6857.  

%%%%%%%%%%%%%%%%%%%%%%%%%%%%%%%%%%%%%%%%%%%%%%%%%%%%%%%%%%%%%%%%%%%%%%%%%%%%%%
% Bibliography
%%%%%%%%%%%%%%%%%%%%%%%%%%%%%%%%%%%%%%%%%%%%%%%%%%%%%%%%%%%%%%%%%%%%%%%%%%%%%%

%\bibliographystyle{aasjournal}
%\bibliography{EXO.bib}

%%%%%%%%%%%%%%%%%%%%%%%%%
%  Figures
%%%%%%%%%%%%%%%%%%%%%%%%%

%\begin{figure*}[h]
%\center
%\includegraphics[trim = 1in 3.5in
%  0.3in .8in,clip, width = 0.8\textwidth]{TrB2_3D_winds}  
%\includegraphics[trim = 1in 3.5in
%  0.3in .8in,clip, width = 0.8\textwidth]{TrB2_3D_density}
%   \llap{\raisebox{0.5cm}{%  move next graphics to top right corner
%      \includegraphics[height=3.cm]{OrbitsLabel}
%      }}
%\caption{Three-dimensional stellar magnetosphere and wind for TRAPPIST-1 simulated using GJ 3622 magnetogram with an average field strength of 600~G.  The Alfv\'en surface is shown in blue and the orbital plane is colored according to the wind speed (top) and to density normalized to that of the solar wind at 1~AU (bottom).  All the known orbits are plotted.}
%\label{fig:3d}
%\end{figure*}

\end{document}